\documentstyle[13lomcon,cite,epsfig,psfig]{article}

\begin{document}

\title{NON-STANDARD NEUTRINO PHYSICS PROBED BY TOKAI-TO-KAMIOKA-KOREA TWO-DETECTOR COMPLEX}

%
%
%
%

\author{ Nei Cipriano Ribeiro$^1$ \footnote{e-mail: ncipriano@fis.puc-rio.br},
Takaaki Kajita$^2$ \footnote{e-mail: kajita@icrr.u-tokyo.ac.jp}, Pyungwon Ko$^3$ \footnote{e-mail: pko@kias.re.kr},
Hisakazu Minakata$^4$ \footnote{e-mail: minakata@phys.metro-u.ac.jp}, Shoei Nakayama$^2$ \footnote{e-mail:
shoei@suketto.icrr.u-tokyo.ac.jp}, Hiroshi Nunokawa$^1$ \footnote{e-mail: nunokawa@fis.puc-rio.br} }

\address{$^1$Departamento de F\'{\i}sica, Pontif{\'\i}cia Universidade Cat{\'o}lica
do Rio de Janeiro, C. P. 38071, 22452-970, Rio de Janeiro, Brazil \\
$^2$Research Center for Cosmic Neutrinos, 
Institute for Cosmic Ray Research, University of Tokyo, Kashiwa, Chiba 277-8582, Japan \\
$^3$School of Physics, Korea Institute for Advanced Study, Seoul 130-722, Korea \\
$^4$Department of Physics, Tokyo Metropolitan University, Hachioji, Tokyo 192-0397, Japan  }


\maketitle

\abstracts{ The discovery potentials of non-standard physics (NSP) 
which might be possessed by neutrinos are examined by 
taking a concrete setting of  
Tokai-to-Kamioka-Korea (T2KK) two detector complex which 
receives neutrino superbeam from J-PARC. 
We restrict ourselves into $\nu_{\mu}$ and $\bar{\nu}_{\mu}$ 
disappearance measurement.  
We describe here only the non-standard interactions (NSI) of 
neutrinos with matter and the quantum decoherence.  
It is shown in some favorable cases T2KK can significantly 
improve the current bounds on NSP. 
For NSI, for example,  $ \varepsilon_{\mu\tau} < 0.03$, 
which is a factor 5 severer than the current one. 
}

\section{Introduction}

The primary objective of the future neutrino oscillation experiments 
is of course to determine the remaining lepton mixing parameters, 
most notably CP violating phase and the neutrino mass hierarchy. 
Nonetheless, it is highly desirable that such facilities possesses 
additional physics capabilities such as exploring possible 
non-standard interactions (NSI) of neutrinos with matter. 
It will give us a great chance of discovering or constraining the 
extremely interesting new physics beyond neutrino mass 
incorporated Standard Model. 
Such additional capabilities are highly desirable because such 
projects would inevitably be rather costly, and it would become 
the necessity if a smoking gun evidence of new physics beyond 
the Standard Model is discovered in $\sim$TeV range.

Some of the present authors have proposed Tokai-to-Kamioka-Korea (T2KK) 
identical two detector complex which receives neutrino superbeam 
from J-PARC as a concrete setting for measuring CP violation and 
determining the mass hierarchy \cite{T2KK1st,T2KK2nd}. 
In this manuscript we report, based on \cite{NSP-T2KK}, discovery reach 
to the possible non-standard interactions of neutrinos and the 
quantum decoherence by the T2KK setting. 
See \cite{NSP-T2KK} for the sensitivities to the Lorentz invariance 
violation as well as the cases which are not treated in this report.

\section{Non-Standard Interactions (NSI) of Neutrinos}

\subsection{NSI; General feature}
\label{NSI-general}

It has been suggested \cite{wolfenstein,grossmann} that neutrinos 
might have non-standard interactions (NSI) which reflect physics 
outside Standard Model of electroweak interactions. 
The possibility of exploring physics beyond the neutrino mass 
incorporated Standard Model is so charming that the sensitivity 
reach of NSI would be one of the most important targets in the 
ongoing as well as future neutrino experiments.  
The latter include neutrino superbeam experiments, reactor 
$\theta_{13}$ experiments \cite{KLOS}, beta beam, and 
neutrino factory \cite{NSI-nufact}. 
See these references for numerous other references on hunting NSI. 
In this sense it is natural to investigate sensitivity reach of NSI by T2KK. 

As a first step we examine the sensitivity to NSI by using $\nu_{\mu}$ 
and $\bar{\nu}_{\mu}$  disappearance modes of T2KK. 
We of course make a comparison between discovery potentials of 
T2KK and Kamioka only (T2K II \cite{T2K}) as well as Korea only settings. 
Our primary concern, however, is not to propose to use NSI sensitivity 
as a criterion of which setting is the best, but rather to understand 
how the sensitivity to NSI in T2KK is determined. 
(The real decision between various settings would require 
many other considerations.)

As is now popular, the effects of NSI are parametrized in a model 
independent way by $\varepsilon_{\alpha \beta}$ parameters 
($\alpha, \beta = e, \mu, \tau$) in the matter sensitive term in 
the effective Hamiltonian in the flavor basis, 
$H^{\rm eff}_{\alpha \beta} = 
a ( \delta_{\alpha e} \delta_{\beta e} + \varepsilon_{\alpha \beta}) $,
where $a \equiv \sqrt{ 2 } G_F N_e $ with 
$G_F$ being the Fermi constant and $N_e $ electron number 
density in the earth.
The existing constraints on $\varepsilon_{\alpha\beta}$ are worked 
out in \cite{constraint}.  

When we restrict ourselves into the disappearance channel 
we can safely truncate the system into the $2 \times 2$ subsystem 
\cite{NSP-T2KK} as 
\begin{eqnarray}
  \label{eq:evol}
i {d\over dt} \left[ \begin{array}{c} 
                   \nu_\mu \\ \nu_\tau 
                   \end{array}  \right]
 = \left[ U \left( \begin{array}{cc}
                   0  & 0  \\
                   0   & \frac{ \Delta m^2_{32} }{2 E} 
                   \end{array} \right) 
            U^{\dagger} +  a \left( \begin{array}{cc}
            0 & \varepsilon_{\mu\tau} \\
             \varepsilon_{\mu\tau}^* & \varepsilon_{\tau\tau} - \varepsilon_{\mu\mu} 
                   \end{array} 
                   \right) \right] ~
\left[ \begin{array}{c} 
                    \nu_\mu \\ \nu_\tau 
                   \end{array}  \right], 
\end{eqnarray} 
where $U$ is the flavor mixing matrix and $a \equiv \sqrt{2} G_F N_e$. 
Because of the form of the 2-2 element of the NSI term in the 
Hamiltonian, we set 
$\varepsilon_{\mu\mu}=0$ and simply discuss the constraint on 
$\varepsilon_{\tau \tau}$ and $\varepsilon_{\mu \tau}$.

\subsection{Sensitivity reach to NSI  }
\label{NSI-sensitivity}

We describe analysis results by skipping the details of the procedure 
by referring the readers \cite{NSP-T2KK} for it. 
The input values $\varepsilon_{\mu \tau}$ and 
$\varepsilon_{\tau\tau} $ 
are taken to be vanishing. 
The important point in correctly estimating the sensitivities is to 
marginalize over the lepton mixing parameters, in particular, 
$\Delta m^2_{32}$ and $\theta_{23}$.

\begin{figure}[t]
\vglue -0.5cm
\centering
\includegraphics[width=9.4cm]{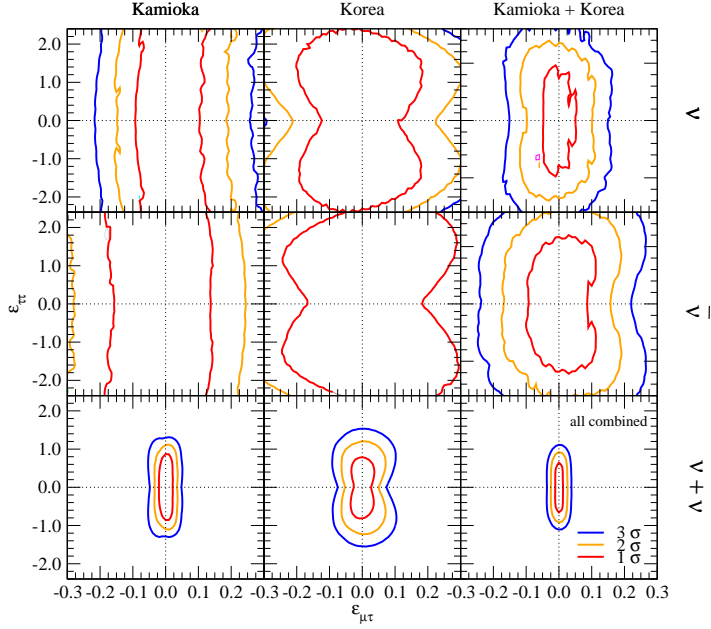}
\vglue -0.1cm
\caption{ The synergy between different detectors and $\nu$ and 
$\bar{\nu}$ running. 
The red, the yellow and the blue lines are for the regions allowed at 
1$\sigma$, 2$\sigma$, and 3$\sigma$ CL (2 DOF), respectively. 
No systematic error is taken into account and 
the number of energy bins considered were 36 from 0.2 to 2.0 GeV. 
The top, the middle, and the bottom panels are for 
4 years neutrino running, 4 years of anti-neutrino running, 
and both neutrino-anti-neutrino combined, respectively. 
The left, the middle and the right panels are the constraints obtained 
by Kamioka detector, by Korea detector (each 0.54 Mton fiducial mass), 
and by both detectors combined, respectively. 
Note that the last one is {\em not} identical with the T2KK setting 
defined which is defined with each fiducial mass of 0.27 Mton, and 
whose sensitivities are presented in Fig.~\ref{NSI-sensitivity}. 
}
\label{synergy}
\end{figure}

To understand competition and synergy between the detectors in 
Kamioka and in Korea, and in particular, between the neutrino 
and the anti-neutrino channels we present Fig.~\ref{synergy}. 
We see from the figure that the Kamioka detector is more sensitive to 
NSI than the Korean detector, probably because of the higher event rate 
by a factor of $\simeq10$. 
The synergy between the neutrino and the anti-neutrino channels 
is striking; 
Neither neutrino only nor anti-neutrino only measurement 
has sensitivity comparable to that of $\nu$ and $\bar{\nu}$ combined.

We present in Fig.~\ref{NSI-sensitivity} the sensitivity to NSI by 
T2KK and its dependence on $\theta_{23}$. 
The approximate
2 $\sigma$ CL (2 DOF) sensitivities of the
Kamioka-Korea setup  for 
$\sin^2 \theta = 0.45 ~(\sin^2 \theta = 0.5)$ are: 
$| \varepsilon_{\mu\tau} | <  0.03 ~(0.03)$ and 
$| \varepsilon_{\tau\tau}|  < 0.3 ~(1.2) $. 
Here, we neglected a barely allowed region near $|\epsilon_{\tau\tau}| = 2.3$,
which is already excluded by the current data. 
Notice that T2KK has potential of (almost) eliminating the island regions. 
The disparity between the sensitivities to $\varepsilon_{\mu \tau}$ and 
$\varepsilon_{\tau\tau} $ can be understood 
by using the analytic formula as discussed in \cite{NSP-T2KK}. 
The figure also contain the comparison between discovery reach 
of NSI by the Kamioka-only setting,  the Korea-only setting, and 
T2KK. 
%

\begin{figure}[t]
\vglue -1.0cm
\centering
\includegraphics[width=8.4cm]{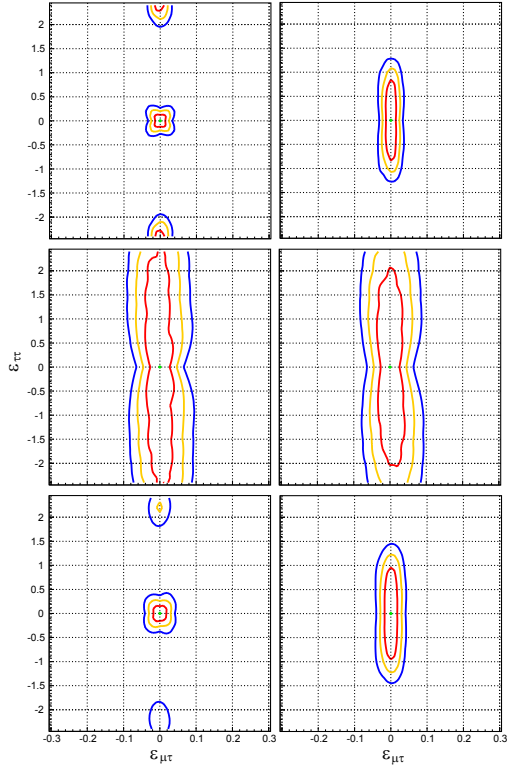}
\vglue -0.8cm
\caption{
The allowed regions in
$\varepsilon_{\mu\tau} - \varepsilon_{\tau\tau}$ space for 4 years neutrino
and 4 years anti-neutrino running.
The upper, the middle, and the bottom three panels are for
the Kamioka-only setting, the Korea-only setting, and the
Kamioka-Korea setting,  respectively.
The left and the right panels are for cases with
$\sin^2 \theta \equiv \sin^2 \theta_{23} = 0.45$ and 0.5, respectively.
The red, the yellow, and the blue lines indicate the allowed regions
at 1$\sigma$, 2$\sigma$, and 3$\sigma$ CL, respectively.
The input value of $\Delta m^2_{32}$ is taken as $2.5\times 10^{-3}$ eV$^2$. 
The figure is taken from \cite{NSP-T2KK}. 
}
\label{NSI-sensitivity}
\end{figure}

\section{Quantum Decoherence}
\label{QD}


Though there is really no plausible candidate mechanism for 
quantum decoherence, people talk about it mainly because 
it can be one of the alternative models for ``neutrino deficit'',  
namely, a rival of the neutrino oscillation. 
It is well known that quantum decoherence modifies the neutrino 
oscillation probabilities. 
The two-level system in vacuum in the presence of quantum decoherence 
can be solved to give the $\nu_{\mu}$ survival probability 
\cite{QD-bari1,benatti-floreanini}:
\begin{eqnarray}
P(\nu_{\mu} \rightarrow \nu_{\mu}) = 
1 -  \frac{1}{2} \sin^2 2 \theta 
\left[ 1 - e^{- \gamma (E) L}
\cos \left( \frac{ \Delta m^2_{32} L} { 2 E} \right) 
\right], 
\label{Pmumu-QD}
\end{eqnarray}
with $\gamma (E) > 0$, the parameter which controls the strength of 
decoherence effect.

The most stringent constraints on decoherence obtained to date 
are by atmospheric neutrino observation 
($\gamma = \gamma_{0} (E/GeV)^2 < 0.9 \times 10^{-27}$ GeV, 
energy-independent $\gamma < 2.3 \times 10^{-23}$ GeV), 
\cite{QD-bari1}, and solar and KamLAND experiments 
($\gamma = \gamma_{0} (E/GeV)^{-1} < 0.8 \times 10^{-26}$ GeV) 
\cite{QD-bari2}. 
(A particular underlying mechanism for decoherence, 
if any, may have some characteristic energy dependence.)  
Yet, such study is worth pursuing in various experiments and in 
varying energy regions because of different systematic errors, 
and for unknown energy dependence of $\gamma$. 
That was our motivation for investigating the sensitivity to 
quantum decoherence achievable in T2KK.

\begin{figure}[t]
\centering
\includegraphics[width=5.5cm]{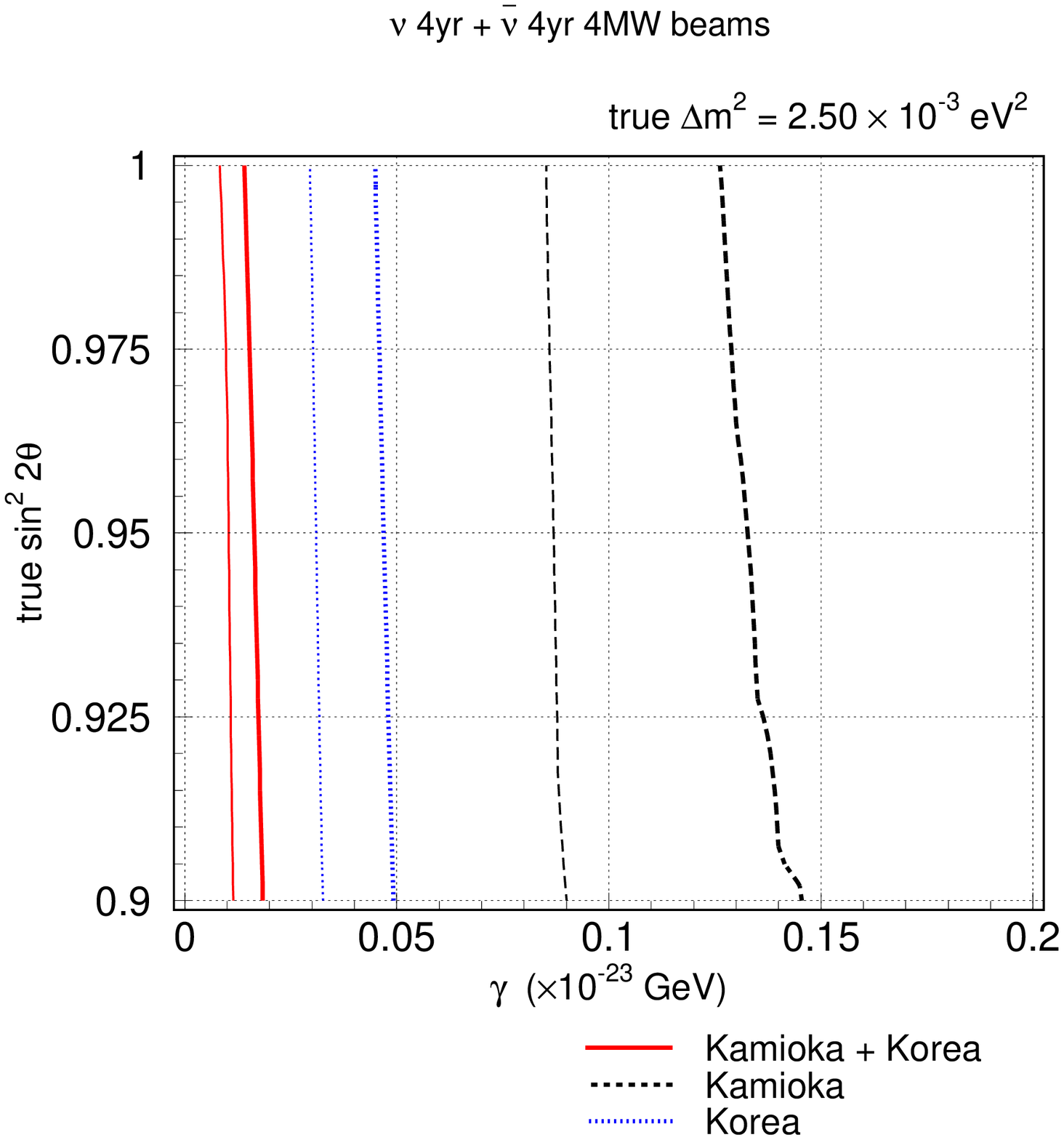}
\includegraphics[width=5.5cm]{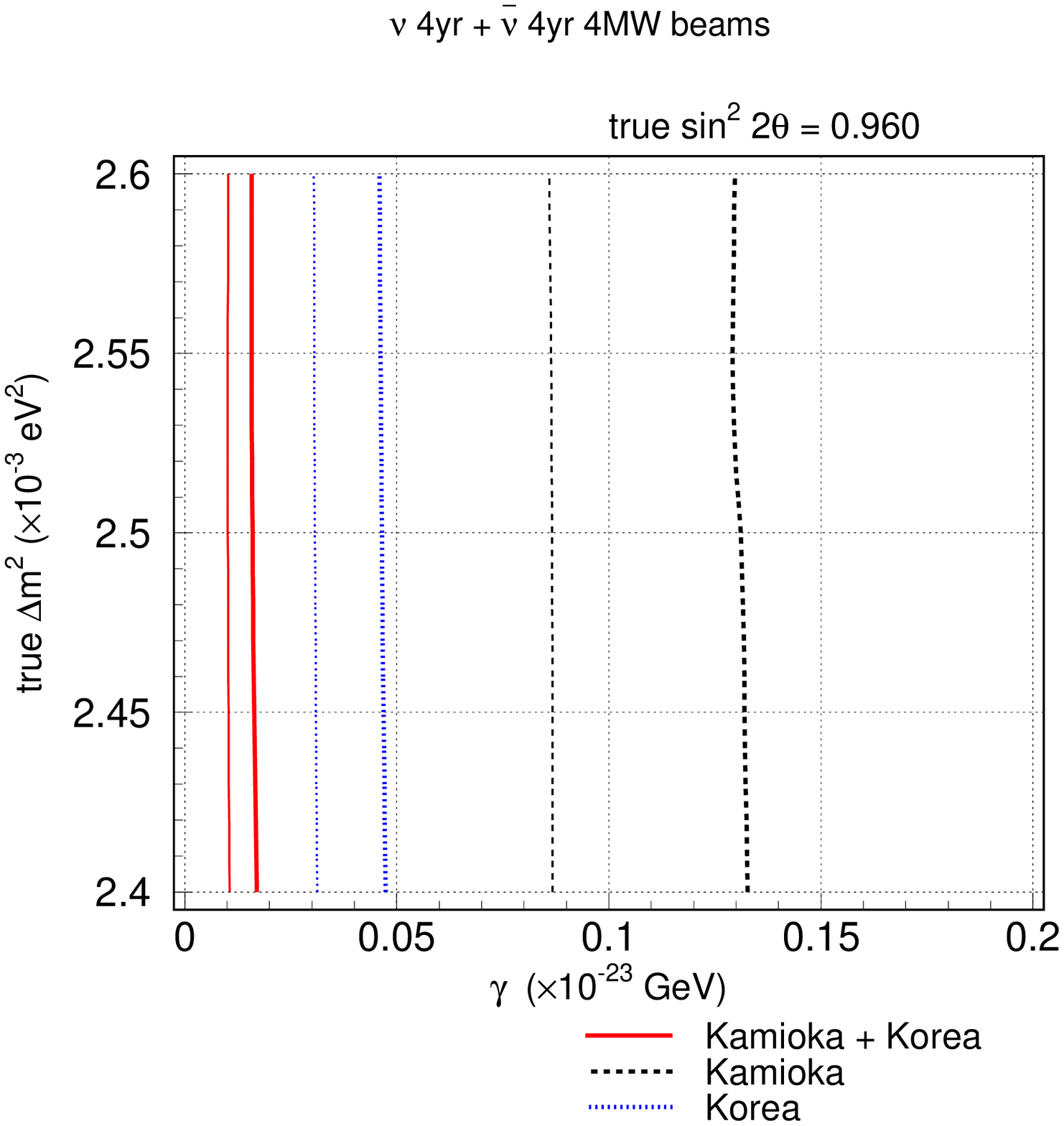}
\caption{ The region of allowed values of $\gamma$ as a function of 
$\sin^2 2\theta \equiv \sin^2 \theta_{23}$ (left panel) and 
$\Delta m^2 \equiv \Delta m^2_{32}$  (right panel). 
The case of energy independence of $\gamma$. 
The red solid lines are for Kamioka-Korea setting with each 0.27 Mton detector, 
while the dashed black (dotted blue) lines are for Kamioka (Korea) 
only setting with 0.54 Mton detector. 
The thick and the thin lines are for 99\% and 90\% CL (1 DOF), 
respectively. 
4 years of neutrino plus 4 years of anti-neutrino running are assumed.  
The normal mass hierarchy is assumed. 
The other input values of the parameters: 
$\Delta m_{31}^2 = +2.5\times 10^{-3}$ eV$^2$, 
$\sin ^2 \theta_{23}$=0.5. 
$\Delta m_{21}^2 = 8\times 10^{-5}$ eV$^2$ and 
$\sin^2{\theta_{12}}$=0.31. 
}
\label{QD-sensitivity;fig}
\end{figure}

In Fig.~\ref{QD-sensitivity;fig} presented is the allowed region of 
the decoherence parameter $\gamma$ as a function of true values 
of $\sin^2 2\theta_{23}$ (left panel) and $\Delta m^2$ (right panel). 
This is the case of energy independent $\gamma$. 
In this case T2KK can improve the current bound on decoherence 
by a factor of 3.  
It is also obvious that the sensitivity to decoherence reachable by 
the T2KK setting far exceeds those of Kamioka-only setting, 
though the sensitivity by Korea-only setting is not so bad.

For cases with alternative energy dependences of $\gamma$ 
and for other additional non-standard physics, see \cite{NSP-T2KK}. 
Most notably, more than 3 orders of magnitude improvement is 
expected in Lorentz-CPT violating parameter.

\section{Conclusion}

In searching for additional physics potential of the Kamioka-Korea 
two-detector setting which receives an intense neutrino beam from 
J-PARC, we have investigated its sensitivities to 
non-standard physics of neutrinos. 
It was shown that T2KK can significantly improve the current bounds 
on quantum decoherence and NSI in some favorable cases.

\section*{Acknowledgments}

H.M. would like to thank the organizers of 13th Lomonosov conference 
for their kind invitation. 
He was supported in part by KAKENHI, Grant-in-Aid for 
Scientific Research, No 19340062, Japan Society for the 
Promotion of Science.

\end{document}